\documentclass[11pt,twoside,letterpaper]{article} %% The same for {book}
\usepackage{times,fancyhdr}
\usepackage[dvips]{graphicx}

\usepackage{epsfig}
\usepackage{amssymb}
\usepackage{amsmath}
\usepackage{amsfonts}
\usepackage{amsthm,amscd}
\usepackage{amsbsy}
\usepackage{latexsym}
\usepackage{bm}
\usepackage{url} 			%% Nicely format and linebreak URLs in the bibliography (and elsewhere).
\usepackage{layout}
\usepackage{pslatex}
\usepackage{cite}
\usepackage{fleqn} 		% displayed formulas flush left (default is centered).
\usepackage{makeidx}
\makeindex 						% Creates index at the end of the book (with makeidx).
\usepackage{layout} 	% To see the current values of these dimensions, use the layout package, 
\usepackage{epstopdf}	% Automatically converts EPS files to encapsulated PDF files (using ghostscript).
\usepackage{color}
\usepackage{hyperref}
\usepackage[latin1]{inputenc}
\usepackage[T1]{fontenc}
\usepackage{poligraf} %% Color separation
\usepackage[letter,cam,center]{crop} %% Printing crop-marks
\usepackage{type1cm} 	%% Use scalable, PostScript Type 1 versions of the Computer Modern fonts.
\usepackage{courier}	%% Replace the standard Computer Modern Typewriter font LaTeX uses
\usepackage{lscape} 	% for landscape section

\makeatletter 
\def\cleardoublepage{\clearpage\if@twoside \ifodd\c@page\else% 
    \hbox{}% 
    \thispagestyle{empty}%
    \newpage% 
    \if@twocolumn\hbox{}\newpage\fi\fi\fi} 
\makeatother

\def\figurename{Figure}
\makeatletter
\renewcommand{\fnum@figure}[1]{\figurename~\thefigure.}
\makeatother

\def\tablename{Table}
\makeatletter
\renewcommand{\fnum@table}[1]{\tablename~\thetable.}
\makeatother

\def\be{\begin{equation}}
\def\ee{\end{equation}}
\def\bea{\begin{eqnarray}}
\def\eea{\end{eqnarray}}

\begin{document}
\title{
{\begin{flushleft}
\vskip 0.45in
{\normalsize\bfseries\textit{Chapter~1}}
\end{flushleft}
\vskip 0.45in
\bfseries\scshape Quadrupolar metrics}}
\author{\bfseries\itshape Hernando Quevedo \thanks{E-mail address: quevedo@nucleares.unam.mx}\\
Instituto de Ciencias Nucleares, \\
 Universidad Nacional Aut\'onoma de M\'exico, \\
AP 70543, M\'exico, DF 04510, Mexico \\
Dipartimento di Fisica and ICRA, \\
Universit\`a di Roma ``La Sapienza'', I-00185 Roma, Italy}

\date{}

\maketitle
\thispagestyle{empty}
\setcounter{page}{1}
% ------- [First Page Running Head] - place it immediately after title! ------
\thispagestyle{fancy}
\fancyhead{}
\fancyhead[L]{In: Book Title \\ 
Editor: Editor Name, pp. {\thepage-\pageref{lastpage-01}}} % needs \label{lastpage-01} on the last page.
\fancyhead[R]{ISBN 0000000000  \\
\copyright~2007 Nova Science Publishers, Inc.}
\fancyfoot{}
\renewcommand{\headrulewidth}{0pt}
%------------------------------------------------------------------------------

\begin{abstract}
We review the problem of describing the gravitational field of compact stars in general relativity. We focus on 
the deviations from spherical symmetry which are expected to be due to rotation and to the natural deformations of mass distributions.
We assume that the relativistic quadrupole moment takes into account these deviations, and consider the class of axisymmetric static 
and stationary  quadrupolar metrics which satisfy  Einstein's equations in empty space and in the presence of matter represented
by a perfect fluid. We formulate  the physical conditions that  must be satisfied for a particular spacetime metric to  
describe the gravitational field of compact stars. We present a brief review of the main static and axisymmetric exact solutions of Einstein's 
vacuum equations, satisfying all  the physical conditions. We discuss how to derive particular stationary and axisymmetric solutions with quadrupolar properties
by using the solution generating techniques which correspond either to Lie symmetries and  B\"ackund transformations of the Ernst equations or to the inverse scattering method applied to  
Einstein's equations. As for interior solutions,  we argue that it is necessary to apply alternative methods to obtain 
physically meaningful solutions, and review a method which allows us to generate interior perfect-fluid solutions.

\end{abstract}

\vspace{.08in} \noindent \textbf{Keywords:} Neutron stars, exact solutions

% ------------ [Running Heads - for odd and even pages] - please insert it only on page 2!
\pagestyle{fancy}
\fancyhead{}
\fancyhead[EC]{Hernando Quevedo}
\fancyhead[EL,OR]{\thepage}
\fancyhead[OC]{Quadrupolar metrics}
\fancyfoot{}
\renewcommand\headrulewidth{0.5pt} 
%------------------------------------------------------------------------------

\section{Introduction}
\label{sec:int}

One of the most important solutions of Einstein's equations is the stationary axisymmetric Kerr solution which 
in Boyer-Lindquist coordinates reads \cite{kerr63}
\be
 ds^2=\frac{(\Delta -a^2\sin^2\theta)}{\Sigma
}dt^2-\frac{2a\sin^2\theta (r^2+a^2-\Delta)}{\Sigma}dtd\varphi \nonumber
\ee
\be
\label{kerr}
\qquad - \left[\frac{(r^2+a^2)^2-\Delta
a^2\sin^2\theta}{\Sigma}\right]\sin^2\theta
d\varphi^2-\frac{\Sigma}{\Delta}dr^2-\Sigma d\theta^2, 
\ee
where 
\be
\Delta = r^2-2m r+a^2 \ ,\quad \Sigma =r^2+a^2\cos^2\theta\ .
\ee
This solution describes the exterior gravitational field of a mass $m$ with specific angular momentum $a=J/m$. 
It is asymptotically flat and reduces to the Minkowski metric in the limit $m=0$ and $a=0$, and to the
Schwarzschild metric in the limit $a=0$. The Kerr spacetime is characterized by the presence of a curvature singularity determined
by the equation 
\be
r^2 + a^2 \cos^2\theta = 0
\ee
which corresponds to a ring located on the equatorial plane $\theta = \pi/2$. This ring singularity, however, cannot be observed from 
outside because it is covered by a horizon located on a sphere of radius
\be
r_h = m+  \sqrt{m^2-a^2} \ .
\ee
Since no information can be extracted from behind the horizon, an external observer will never be aware about the existence of the ring singularity.
In this sense, the singularity can be considered as non-existing for observers located outside the horizon. The Kerr spacetime can be therefore interpreted
as describing the exterior gravitational field of a rotating black hole. Furthermore, the black hole uniqueness theorems \cite{heus96} state that the Kerr spacetime
is the most general vacuum solution that corresponds to a black hole. In other words, to  describe a black hole, we only need two parameters, namely, 
mass and angular momentum. 

In the case $a^2>m^2$, no  horizon exists 
and the ring singularity becomes naked. However, several  studies \cite{def78,cal79,rud98} show 
that in realistic situations, where astrophysical objects are surrounded by accretion disks, a
Kerr naked singularity is an unstable configuration that 
rapidly decays into a Kerr black hole. Furthermore, it now seems to be well established that  in generic situations a 
gravitational collapse cannot lead to the formation of a Kerr naked singularity.
These results seem to indicate that rotating Kerr naked singularities do not exist in Nature.
Again, these results  corroborate that the Kerr spacetime describes rotating black holes. 

From an astrophysical point of view, black holes belong to the class of compact objects which include also neutron stars and white dwarfs. 
The question arises whether the Kerr metric can also  be used to describe the exterior gravitational field of neutron stars and white dwarfs. 
To try to answer this question, let us recall that from the point of view of general relativity, the gravitational field of a compact source should be described by a complete Riemannian  differential manifold, i.e., it should include an exterior metric and an interior metric as well. Let us suppose for a moment 
that the Kerr spacetime describes the exterior field of all compact objects, and consider the interior counterpart. 
In the case of black holes, in which 
the matter content of the original star has collapsed to form a curvature singularity, we argue that it is not possible to find 
the interior counterpart within the framework of classical general relativity. Indeed, since all the information about the internal structure
of a black hole is located inside the singularity, where the classical theory is not valid any more, we should apply an alternative theory 
that must take into account the effects of gravity under extreme pressures and densities, as intuitively expected at the singularities. Such
a theory could be quantum gravity which, in the best case, is still under construction. This argument implies that the quantum interior counterpart
of the Kerr metric is well beyond our reach in the short term. 

Consider now the interior field of neutron stars and white dwarfs. An interior 
metric should describe an equilibrium structure, probably a fluid, bounded by a surface of zero pressure and matched across this surface to the exterior 
Kerr metric. The search for such an interior solution has been conducted for over 50 years, and not even a single physical meaningful solution 
has been found to date. Many arguments can be found to explain this negative result, especially, regarding the relatively simple models used 
to describe the internal structure of such compact stars. Nevertheless, if we consider a more elaborated internal model, the mathematical complexity 
of the field equations and the matching conditions usually increases as well, implying that the possibility of solving the problem decreases.
This is probably the reason why the search for physically meaningful interior solutions has not been very successful. 
In our opinion, the simplest solution to this problem is to assume that the Kerr metric does not describe the exterior field of rotating compact 
objects, but black holes.  This is exactly the working hypothesis we will assume henceforth.

The question arises: What metric should we use to describe the exterior field of neutron stars and white dwarfs?  The black hole uniqueness theorems
\cite{heus96} sheds some light on how to look for an answer to this question. In fact, black holes are described by only the mass $m$ and the angular momentum $J$.
From the point of view of the multipole structure of exact vacuum solutions (for a review see, for instance, \cite{quev90}), this is equivalent to saying
 that only the lowest multipoles are present in black holes, namely, the mass monopole $m$ and the angular-momentum dipole $J$.  Then, it seems reasonable 
to include higher moments in order to describe the exterior field of compact objects, other than black holes. The simplest choice to begin with is the 
mass quadrupole. Consequently, we assume in this work that to describe the exterior field of neutron stars and white dwarfs, we need a vacuum metric 
with three physical parameters, namely, mass $m$, angular momentum $J$, and quadrupole $q$\footnote{Of course, one could also include higher multipoles like the mass octupole, the angular-momentum quadrupole, etc. However, we limit ourselves here to the lowest non-ignorable multipole which is the mass quadrupole.}. 
From a physical point of view, it is also reasonable to 
consider the mass quadrupole as an additional parameter, because it represents the natural deviations of  a mass distribution from the ideal spherical symmetry.
In other words, we assume that in the case of neutron stars an white dwarfs, it is not possible to neglect the gravitational field generated by the quadrupole,
whereas in the case of black holes, the uniqueness theorems prove that the quadrupole is zero. 

On the other hand, since the uniqueness theorems are valid only in the case of mass and angular momentum, with the Kerr metric as the only exact solution, 
 there must exist several exact solutions with mass, angular momentum and quadrupole.  
The main goal of this work is to present a review and a brief description of the main exact vacuum solutions of Einstein equations with mass quadrupole. 

This paper is organized as follows. In Sec. \ref{sec:stat}, we focus on static gravitational sources. We present the field equations and describe the most
important properties that a metric should satisfy in order to describe the exterior field of a compact source. We present the explicit form of the 
metrics that, to our knowledge, have been used in general relativity to describe the field of static mass distributions. In Sec. \ref{sec:star}, we study the 
rotating generalizations of the quadrupolar metrics. Then, in Sec. \ref{sec:interior}, we describe the uninspiring situation in the case of interior solutions. 
Finally, in Sec. \ref{sec:con}, we discuss the situation in general and comment on the open problems regarding the description of the gravitational field of 
neutron stars and white dwarfs. 

%%%%%%%%%%%%%%%%%%%%%%%%%%%%%%%%%%%%%%%%%%%%%%%%%%%%%%%%%
%%%%%%%%%%%%%%%%%%%%%%%%%%%%%%%%%%%%%%%%%%%%%%%%%%%%%%%%%

\section{The gravitational field of compact stars}
\label{sec:cstars}

To describe the exterior gravitational field it is necessary to obtain exact solutions of Einstein's equations in empty space. 
Since Einstein's field equations are in general difficult to handle, especially when the aim is to obtain 
physically meaningful solutions, it is necessary to assume the validity of certain physical conditions about the problem under consideration.
We assume that the gravitational field of compact stars do not change drastically in time so that
stationarity can be adopted. In general, we know from observations that this condition is satisfied in most 
astrophysical objects. Moreover, the assumption  of stationarity does not exclude the possibility of 
rotation which is an important characteristic of all known compact stars. 
 To consider the deviations of the mass distribution from spherical symmetry, we will assume the 
existence of an axis of symmetry which for the sake  of simplicity is supposed to coincide with the axis 
of rotation. Moreover, to take into account the deformations of the mass distribution with respect to 
the axis of symmetry, we will consider only the quadrupole moment.
 
The above assumptions imply that we must focus our analysis on the case of stationary axisymmetric
gravitational fields. The corresponding line element in the case of empty space is known as the 
Weyl-Lewis-Papapetrou \cite{weyl17,lew32,pap66} line element that in cylindrical coordinates can be written as
\cite{solutions}
\be
ds^2 = f(dt-\omega d\varphi)^2 - \frac{1}{f}\left[ e^{2\gamma}(d\rho^2 + dz^2) + \rho^2 d\varphi^2\right] 
\ .
\label{lel}
\ee
Here $t$ and $\varphi$ are the coordinates associated to the Killing vector fields 
$\eta^I = \partial_t$ and $\eta^{II} = \partial_\varphi$
which represent stationarity and axial symmetry, respectively. The functions $f$, $\omega$ and $\gamma$ 
depend on the spatial coordinates $\rho$ and $z$.  A straightforward computation shows that Einstein's 
vacuum field equations reduce in this case to 
\be
f(f_{\rho\rho} + f_{zz} + \rho^{-1} f_\rho) - f_\rho^2 - f_z^2 + \rho^{-2} f^4(\omega_\rho^2+\omega_z^2) = 0 \ ,
\ee
\be
f(\omega_{\rho\rho} + \omega_{zz} - \rho^{-1} \omega_\rho) + 2 f_\rho\omega_\rho + 2 f_z\omega_z = 0 \ ,
\ee
\be
\gamma_\rho = \frac{1}{4} \rho f^{-2}(f_\rho^2 - f_z^2) - \frac{1}{4} \rho^{-1} f^2(\omega_\rho^2 - \omega_z^2)\ ,
\ee
\be
\gamma_z = \frac{1}{2} \rho f^{-2} f_\rho f_z - \frac{1}{2} \rho^{-1} f^2 \omega_\rho \omega_z \ ,
\ee
where $f_\rho = \partial f/\partial \rho$, etc. From this set of equations it follows that $f$ and $\omega$
can be considered as the main metric functions since $\gamma$ can be calculated by quadratures once the main
functions are known.

%%%%%%%%%%%%%%%%%%%%%%%%%%%%%%%%%%%%%%%%%%%%%%%%
%%%%%%%%%%%%%%%%%%%%%%%%%%%%%%%%%%%%%%%%%%%%%%%
\subsection{The Ernst representation}

Although the field equations are apparently highly non-linear and complicated, it turned out that
they are characterized by an internal symmetry related to the Lie transformations of differential 
equations. The investigation of this internal symmetry led to the formulation of modern
solution generating techniques which, when formulated in an appropriate manner, allow us to generate new 
solutions from known ones \cite{diehoe}. 

The starting point for the understanding of the Lie transformations was
the Ernst \cite{ernst} representation of the field equations. Ernst proposed a simple 
Lagrangian density from which the field equations can be derived. Moreover, 
Ernst's Lagrangian can be derived from the Einstein-Hilbert Lagrangian and, in fact, 
similar representations \cite{cnq} can be derived for any gravitational fields with two commuting 
Killing vector fields, for example, for Einstein-Rosen gravitational waves or  inhomogeneous
cosmological Gowdy models. In the case  under consideration, it can be shown that the Einstein-Hilbert 
action can be reduced to the form
\be
{\cal L}_{EH} = 
 \frac{\rho}{2f^2}\left( f_\rho^2 + f_z^2 + \Omega_\rho^2 + \Omega_z ^2\right) + {\hbox{boundary terms}} \ ,
\label{redlag}
\ee
where the new function $\Omega$ is defined by the relationships 
\be
\rho\Omega_\rho = f^2 \omega_z\ , \quad \rho\Omega_z = - f^2 \omega_\rho\ .
\ee
  Neglecting the boundary terms,
the above action can be expressed as
\be
{\cal L}_{EH} = \frac{\rho}{2f^2}\left[ (Df)^2 + (D\Omega)^2\right] = \frac{\rho}{2f^2}(DE) (DE^*)  \ ,
\quad E = f + i \Omega\ , \quad  D=(\partial_\rho,\partial_z) \ ,
\label{ernstlag}
\ee
where $E$ is the Ernst potential which is complex for stationary fields and real in the 
limiting static case $(\Omega=0)$. 
Finally, it can be shown that the main field equations can be obtained 
as 
\be
\frac{\delta {\cal L}_{EH}}{\delta f} = 0\ ,\quad \frac{\delta {\cal L}_{EH}}{\delta \Omega} = 0 
\quad \Leftrightarrow \quad 
 \frac{\delta {\cal L}_{EH}}{\delta E} = 0 \ .
\ee

Clearly, the compactness and simplicity of Ernst's formulation is very useful for investigating 
the field equations. For instance, from (\ref{ernstlag}) it can be seen that the transformation 
$E\rightarrow e^{i\tau} E$, with $\tau=const$, does not affect the Lagrangian and, consequently, the 
field equations. It follows that this ``gauge"  transformation can be used to generate new solutions. 
In fact, it is known that the Kerr-NUT (Newman-Unti-Tamburino) solution can be obtained from the Kerr solution by  applying a ``gauge"
transformation. A further computational advantage of the use of the Ernst representation is that it is coordinate
invariant in the sense that in order to analyze it in a different coordinate system, one only needs to 
calculate the explicit form of the vector operator $D$ in the new coordinates.

Finally, we would like to mention that the Ernst representation reproduces only partially Einstein's equations
because it includes only the main functions $f$ and $\omega$. The metric function $\gamma$  does not appear 
in the reduced Lagrangian (\ref{redlag}) because it turns out to be a cyclic coordinate in the Einstein-Hilbert
Lagrangian which is absorbed by a Legendre transformation. However, it is possible to interpret the reduced Lagrangian 
as determining the action of a generalized harmonic map  from which the equations for the function $\gamma$
can be recovered \cite{hnq09}.

%%%%%%%%%%%%%%%%%%%%%%%%%%%%%%%%%%%%%%%%%%%%%%%%%%%%%%%%%%%%%%%%%
\subsection{Physical conditions}
\label{sec:physcon}

One can find many stationary axisymmetric solutions of Einstein's equations, but not all them are necessarily 
suitable to describe the gravitational field of compact stars. Several physical conditions must be imposed which can be 
described as follows.

\begin{description}

\item {$i)$} The spacetime must be asymptotically flat. This means that far away from the source the gravitational field should be
negiglible small, and can be described approximately by the Minkowski metric. 

\item $ii)$ The spacetime must be elementary flat, i.e., the axis of symmetry must be free of conical singularities. This 
property means that the coordinate $\varphi$ is a well-defined angle coordinate that can be used to represent the 
rotation of the compact star. 

\item 
$iii)$ The spacetime must be free of singularities outside the surface of the star. Curvature singularities can exist 
inside the surface where the vacuum solution is not valid any more and, instead, an interior solution should exist 
that ``covers'' the singularity.

\item  
$iv)$ The spacetime must be free of horizons in order to be in accordance with the black hole uniqueness theorems.

\item
$v)$  The solution must reduce  to the Minkowski metric in the limiting case when the mass monopole vanishes, independently of the
values of the remaining parameters. This condition guarantees that there are no rotations and no deviations from spherical symmetry
without the presence of a physical mass distribution.

\item
$vi)$ The solution must be matched with a physically meaningful interior solution across the surface of the star where the pressure and
the density of the interior configuration should vanish.

\end{description}

The fulfillment of these conditions represents the real challenge for describing the gravitational field of compact stars. Whereas there many 
metrics that can be used to represent the exterior field, the interior counterparts are still  unknown.

%%%%%%%%%%%%%%%%%%%%%%%%%%%%%%%%%%%%%%%%%%%%%%%%%%
\subsection{Prolate spheroidal coordinates}
\label{sec:psc}

Any exact solution of Einstein's equations can be written in many equivalent ways by using different coordinate systems. Nevertheless, some
coordinate systems are specially adapted to the geometric properties of the field configurations. In the case of stationary axisymmetric solutions,
the system of prolate spheroidal coordinates has been used extensively in the scientific literature to represent several important particular solutions. 
In this case, the general line element can be expressed as 
\bea
& & ds^2 = f(dt-\omega d\varphi)^2 \nonumber \\
& - &
\frac{\sigma^2}{f}\left[ e^{2\gamma}(x^2-y^2)\left( \frac{dx^2}{x^2-1} + \frac{dy^2}{1-y^2} \right) 
+ (x^2-1)(1-y^2) d\varphi^2\right] \ ,
\label{lelxystar}
\eea
where all the metric functions depend on $x$ and $y$, only. The simplest way to represent the corresponding field equations in this case is by using 
the complex Ernst potentials
\be
E=f+i\Omega \ , \quad \xi = \frac{1-E}{1+E} \  ,
\ee
where the function $\Omega$ is now determined by the equations
\be
 \sigma (x^2-1)\Omega_x = f^2\omega_y\ ,\quad  \sigma (1-y^2)\Omega_y = - f^2 \omega_x \ . 
\ee
Then, one can show that the main field equations can be represented in a compact and symmetric form  as 
\be 
(\xi \xi^*-1) \left\{[(x^2-1)\xi_x]_x + [(1-y^2)\xi_y]_y\right\} = 2 \xi^*[ (x^2-1)\xi_x^2 + (1-y^2)\xi_y^2]\ ,
\label{eqksixy}
\ee
where the asterisk represents complex conjugation. Notice that in the case of static fields the Ernst potentials 
become real and the above equation generates a linear differential equation for $\psi=(1/2)\ln f$ for which the
general solution can be obtained. Moreover, it is easy 
to see that equation (\ref{eqksixy}) 
is invariant with respect to the transformation $x\leftrightarrow y$. This simple symmetry property can be used to generate new solutions.
Indeed, consider the particular solution 
\be
\xi = \frac{1}{x} \rightarrow \Omega=0 \rightarrow \omega = 0 \rightarrow f= \frac{x-1}{x+1}\ , \quad \gamma = \frac{1}{2}\ln \frac{x^2-1}{x^2-y^2}
\ee
which can be shown to represent the Schwarzschild spacetime. Then, the function $\xi^{-1} = y$ is also an exact solution. Furthermore, if we take the
linear combination $\xi^{-1} = c_1 x+c_2y$, with arbitrary constants $c_1$ and $c_2$, 
 and introduce it into the field equation (\ref{eqksixy}), we obtain the new solution 
\be 
\xi^{-1} = \frac{\sigma}{m} x+i\frac{a}{m} y\ ,\  \sigma =\sqrt{m^2-a^2}\ ,
\ee
which corresponds to the Kerr metric in prolate spheroidal coordinates. The corresponding metric functions 
are 
\begin{eqnarray}
f&=&\frac{c^2x^2+d^2y^2-1}{(cx+1)^2+d^2y^2}\ , \quad 
\omega=2a\frac{(cx+1)(1-y^2)}{c^2x^2+d^2y^2-1}\ , \nonumber\\
\gamma &=&\frac12\ln\left(\frac{c^2x^2+d^2y^2-1}{c^2(x^2-y^2)}\right)\ , \quad
c=\frac{\sigma}{m}\ , \quad d=\frac{a}{m}\ , \quad c^2+d^2=1\ .
\end{eqnarray}

%%%%%%%%%%%%%%%%%%%%%%%%%%%%%%%%%%%%%%%%%%%%%%%%%%%%%%%%%
%%%%%%%%%%%%%%%%%%%%%%%%%%%%%%%%%%%%%%%%%%%%%%%%%%%%%%%%%

\section{Static quadrupolar metrics}
\label{sec:stat} 

The simplest case of a multipolar spacetime is described by the Schwarzschild metric which possesses only the mass monopole. 
Birkhoff's theorem \cite{solutions} guarantees that this metric is unique. Furthermore, from a physical point of view, one expects that 
a dipole moment can be made to vanish by an appropriate coordinate transformation which, in the Newtonian limit, corresponds to locating
the origin of spatial coordinates on the center of mass of the object. The next interesting configuration consists of a mass with quadrupole
moment. In this case, no uniqueness theorem exists and, therefore, we can expect that Einstein's equations permit the existence of several 
solutions describing such a gravitational system. Indeed, several exact solutions are known.

Weyl \cite{weyl17} found the most general static axisymmetric asymptotically flat solution in cylindrical coordinates (\ref{lel})
\be
\ln f = 2 \sum_{n=0}^\infty \frac{a_n}{(\rho^2+z^2)^\frac{n+1}{2}} P_n({\cos\theta}) \ ,
\qquad \cos\theta = \frac{z}{\sqrt{\rho^2+z^2}} \ ,
\label{weylsol}
\ee
where $a_n$ $(n=0,1,...)$ are arbitrary constants, and $P_n(\cos\theta)$ represents the Legendre
polynomials of degree $n$. 
As mentioned above, the metric function $\gamma$ 
can be calculated by quadratures. Then, we obtain \cite{solutions}
\be
\gamma = - \sum_{n,m=0}^\infty \frac{ a_na_m (n+1)(m+1)}{(n+m+2)(\rho^2+z^2)^\frac{n+m+2}{2} }
\left(P_nP_m - P_{n+1}P_{m+1} \right) \ .
\ee
The set of parameters $a_n$ essentially determines the set of mass multipoles $M_n$ as computed by using the Geroch-Hansen definition \cite{ger70a,ger70b,han74}, for instance.
Then, a configuration composed of a mass and a quadrupole can be written as
\be
\frac{1}{2} \ln f = \frac{a_0} { (\rho^2 + z^2)^{1/2} } + \frac{a_2}{  (\rho^2 + z^2)^{3/2} } P_2(\cos\theta) 
\ee
The first term is called the Chazy-Curzon metric \cite{solutions} and describes the field of two particles located along the symmetry axis with a curvature singularity among them, i.e., it corresponds to a strut located along the axis.
 The second term can be considered as representing a quadrupole deformation of the strut. Far away from the source,  the
Chazy-Curzon metric leads to the Newtonian potential of a point particle. One could therefore expect that in the Newtonian limit the second term generates a quadrupole moment. From a physical point of view, one would expect that close to a non-rotating compact star with no quadrupole, the metric is spherically symmetric. We see that the above Weyl metric does not satisfy this condition. We therefore conclude that it cannot be used to describe the exterior field of compact stars.

To our knowledge, Erez and Rosen \cite{erro59} found the first quadrupolar metric which reduces to the Schwarzschild metric in the limit of vanishing quadrupole. In prolate spheroidal coordinates (\ref{lelxystar}) it can be expressed as ($q_2$ is a constant) 
\be
\ln f = \ln \frac{x-1}{x+1} + q_2 (3y^2-1)\left[ \frac{1}{4} (3x^2-1) \ln \frac{x-1}{x+1} + \frac{3}{2} x \right] \ ,
\ee
\bea 
\gamma = && \frac{1}{2} (1+q_2)^2 \ln \frac{x^2-1}{x^2-y^2} - \frac{3}{2} q_2 (1-y^2) \left( x \ln \frac{x-1}{x+1} +2\right)
\nonumber\\
&& +\frac{9}{16} q_2^2 (1-y^2) \bigg[ x^2 + 4 y^2 - 9x^2 y^2 - \frac{4}{3} \nonumber \\
&& +x\left(x^2 + 7 y^2 - 9 x^2y^2 -\frac{5}{3}\right) \ln \frac{x-1}{x+1} \nonumber \\
&&+ \frac{1}{4}(x^2-1)(x^2 +y^2 - 9 x^2y^2 -1) \ln^2 \frac{x-1}{x+1}\bigg] \ .
\eea
In the limiting case $q_2\rightarrow 0$, the Erez-Rosen metric reduces to the Schwarzschild metric, as expected for a compact star.
In general, this solution is asymptotically flat and free of singularities outside the spatial region determined by $x=1$, which in the case of vanishing quadrupole corresponds to the Schwarzschild radius. It also satisfies the condition of elementary flatness. From this point of view, the Erez-Rosen solution satisfies all the conditions to describe the exterior field of a
deformed mass with quadrupole moment. However, no interior solution is known that could be matched with the exterior metric 
on the surface of the body.

Gutsunayev and Manko  \cite{gm85} derived the following  exact static solution ($A_2$ is a constant)
\be
\ln f = \ln \frac{x-1}{x+1} + 2 A_2 \frac{x(x^2-3x^2y^2 + 3 y^2 - y^4)}{(x^2-y^2)^3}\ ,
\ee
\bea
\gamma = && \frac{1}{2} \ln \frac{x^2 -1}{x^2-y^2} + \frac{A_2(1-y^2)}{2(x^2-y^2)^4} [ 3(1-5y^2)(x^2-y^2)^2 \nonumber\\
&& + 8 y^2 (3-5Y^2)(x^2-y^2) +24 y^4 (1-y^2) ] + \frac{A_2^2 (1-y^2)}{8(x^2-y^2)^8}\nonumber\\
&& \times [-12 (1-14y^2 +25y^4)(x^2-y^2)^5 + 3 (3-153 y^2 \nonumber\\
&& + 697 y^4 -675 y^6) (x^2-y^2)^4 \nonumber\\
&& + 32 y^2 (9-105 y^2 +259 y^4 - 171 y^6)(x^2-y^2)^ 3 \nonumber\\
&& + 32 y^4 (45 - 271 y^2 + 451 y^4 -225 y^6)(x^2-y^2)^ 2\nonumber\\
&& + 2304 y^6 (1 - 4 y^2 + 5 y^4 - 2y^6) (x^2-y^2) \nonumber \\
&& + 1152 y^8(1-3y^2+3y^4-y^6)] \ .
\eea
Although at first glance these two solutions look quite different, it is possible to show \cite{quev87}  that if we choose the 
parameters as 
\be 
A_2 = \frac{1}{15} q_2 \ , 
\ee
the quadrupole moment of both metrics coincide, but differences appear at the level of the $2^4$-pole moment.  

A different quadrupolar metric was derived by Hern\'andez-Pastora and Mart\'\i\ \cite{hm94} which is also 
given in prolate spheroidal coordinates:
\be
\ln f = \ln \frac{x-1}{x+1} + \frac{5}{4} B_2 \left\{ \frac{3}{4} [(3x^2-1)(3y^2-1) - 4] \ln \frac{x-1}{x+1} 
-\frac{2x}{x^2 - y^2} + \frac{3}{2} x(3y^2-1) \right\}\  .
\ee
As in the previous cases, the corresponding $\gamma$ function can be calculated by quadratures by using the 
explicit for of $f$ only. The resulting expression is quite complicated. We refer to the original paper
for the explicit expression. In the above solution, the constant parameter $B_2$ essentially determines 
the quadrupole moment of the mass distribution. 

Recently, in \cite{fs16}, the multipole moment structure of the above solutions 
with free parameters $q_2$, $A_2$ and $B_2$ was investigated in detail with the result that 
the Geroch quadrupole moment of all three metrics can be made to coincide by choosing the free parameters
appropriately. On the other hand, it is known that in general relativity, stationary and axisymmetric vacuum spacetimes 
can be completely characterized by their multipolar structure and  if two spacetimes have the same moments,
then they represent essentially the same spacetime. If then follows that the above metrics with free parameters
 $q_2$, $A_2$ and $B_2$ are in fact the same spacetime, if we consider only the quadrupole moment. However, if 
higher moments are taken into account, differences appear that make the three metrics different from a physical 
point of view. 

As can be seen from the above expressions, the explicit form of the known quadrupolar metrics is not simple, usually 
making them  difficult to be analyzed. In a recent work \cite{quev11a}, we proposed an alternative solution
as the simplest generalization of the Schwarzschild solution which contains a quadrupole parameter $q$. In spherical coordinates,
it has the simple and compact expression  
\bea
ds^2 = &&  \left(1-\frac{2m}{r}\right) ^{1+q} dt^2 - \left(1-\frac{2m}{r}\right)  ^{-q} \nonumber  \\
&& \times \left[ \left(1+\frac{m^2\sin^2\theta}{r^2-2mr }\right)^{-q(2+q)} \left(\frac{dr^2}{1-\frac{2m}{r} }+ r^2d\theta^2\right) + r^2 \sin^2\theta d\varphi^2\right] \ .
\label{zv}
\eea
This solution is obtained from the Schwarzschild metric  by applying a Zipoy-Voorhees transformation \cite{zip66,voo70}.
In the literature, for notational reasons this solution is known as the  $\delta-$metric or as the $\gamma-$metric \cite{mala04}.
Instead, we propose to use the term quadrupole metric ($q-$metric) to emphasize the role of the parameter $q$ which determines the quadrupole moment. Indeed, a straightforward computation of the Geroch multipole moments leads to a monopole $ M_0= (1+q)m$ and a quadrupole $M_2 = -\frac{m^3}{3}q(1+q)(2+q)$. If $q=0$, we obtain the limiting case of the Schwarzschild metric. Moreover, the free parameters $m$ and $q$ can be chosen in such a way that the quadrupole moment $M_2$ is negative (oblate objects) or positive 
(prolate objects). Furthermore, one can easily show that this solution satisfies all physical conditions mentioned in the previous 
section for exterior solutions. This implies that it can be used to describe the exterior gravitational field of static compact stars. 
A detailed analysis of the circular motion of test particles around a compact object described by the $q-$metric shows that the presence 
of the quadrupole parameter can drastically change the physical behavior of test particles, and the obtained effects corroborate the interpretation of
$q$ as determining the deviation of the mass distribution from spherical symmetry \cite{bggqt16}.  

%%%%%%%%%%%%%%%%%%%%%%%%%%%%%%%%%%%%%%%%%%%%%%%%%%%%%%%%%
%%%%%%%%%%%%%%%%%%%%%%%%%%%%%%%%%%%%%%%%%%%%%%%%%%%%%%%%%

\section{Stationary quadrupolar metrics}
\label{sec:star} 

All the solutions presented in the previous section do not take into account an important characteristic of compact objects, namely, the rotation. Realistic 
exact solutions should contain at least one additional parameter that could be interpreted as rotation. In terms of multipole moments, this means that the 
angular-momentum dipole should be nonzero. The first exact solution with a non-trivial angular-momentum dipole was discovered by Kerr in 1963. Soon after, Ernst
proposed a general representation for stationary and axisymmetric vacuum and electrovacuum spacetimes that allowed researchers in this field to derive 
a new type of internal symmetries of the field equations. As a result, some solution generating techniques \cite{solutions} were developed whose main 
objective is to generate new solutions from known ones.

The first generating methods such as the Kerr-Schild Ansatz, the complex Newman-Janis Ansatz, 
and the Hamilton-Jacobi separability procedure were limited to generate only the (charged) 
Kerr-NUT (Newman-Unti-Tamburino) class of stationary solutions. Nevertheless,
the simple and compact Ernst representation was used by Tomimatsu and Sato and Yamazaki and Hori to 
find exact solutions with a particular functional dependence for the Ernst potential. Furthermore, Ernst
developed two generating methods that were generalized by Kinnersley \cite{solutions}.

All the early methods were based on particular symmetries of the field equations. 
The discovery of Lie symmetries of the Ernst representation in the
late seventies determined the starting point for the development of modern solution generating
techniques. All the symmetry transformations of the field equations involve in general an 
infinite dimensional group of transformations. One of the main difficulties was to isolate
only those transformations that preserve asymptotic flatness and do not generate unphysical curvature 
singularities {\it a priori}. Finally, Hoenselaers, Kinnersley and Xanthopoulos found
subgroups of the Geroch group which preserve asymptotic flatness and can easily be
extrapolated by purely algebraic methods.

Particular cases of B\"acklund transformations of the Ernst equations were found by Harrison 
and Neugebauer. B\"acklund transformations were first used to generate asymptotically flat
solutions, using the Minkowski metric as seed solution. In general, it can be shown that
the generated solution is asymptotically flat, if this is also a property of the seed solution. 

A different method was proposed by Belinsky and Zakharov in which the nonlinear Einstein field
equations are represented as a linear eigenvalue problem which can be solved by means of the inverse
scattering method. This method allows one to generate solitonic solutions, one of which 
corresponds to the Kerr-NUT solution. 

All the above methods imply several detailed procedures with quite complicated calculations. 
A particularly simple and different method was developed by Sibgatullin \cite{sib91} in which 
only the value of the Ernst potential on the axis of symmetry is required in order to calculate 
the general form of the potential from which the corresponding metric can be calculated. 
Suppose that the Ernst potential in cylindrical coordinates is given as an arbitrary 
function $e(z)$ on the axis $\rho=0$. Then, the Ernst potential for the entire spacetime can be calculated as
\be
E(\rho,z) = \frac{1}{\pi} \int_{-1}^{+1} \frac{e(\xi) \mu(\sigma) }{\sqrt{1-\sigma^2} } d\sigma\ ,
\ee
where $\xi=z + i \rho\sigma$ and the unknown function $\mu(\sigma)$ satisfies the singular integral equation
\be
\int_{-1}^{+1} \frac{ \mu(\sigma)[e(\xi) +   e^* (\eta^*) ]}{ (\sigma-\tau)\sqrt{1-\sigma^2} } d\sigma = 0 \ ,
\ee
and the normalizing condition 
\be
\int_{-1}^{+1} \frac{\mu(\sigma)}{\sqrt{1-\sigma^2}} d\sigma = \pi\ ,
\ee
where $\eta = z+ i \rho \tau$, and the asterisk represents complex conjugation. This method has been used to generate several stationary 
and axisymmetric solutions \cite{mms00,prs06} which satisfy all the conditions to describe the exterior field of neutron stars and 
are in accordance with a series of observations. These solutions are characterized by a finite number of parameters which are interpreted
in terms of multipoles. For instance, the most general  solution of this class has six  parameters and is determined on the axis
by the Ernst potential \cite{prs06}
\be
e(z) = \frac{z^3 - (m+ia) z^2 - k z +i s}{ z^3 + (m-ia) z^2 -k z + is}\ ,
\ee
which contains four parameters. An additional function corresponding to the electromagnetic potential on the axis contains the two remaining parameters.
In the case of vanishing electromagnetic field $(s=0)$ and rotation $(a=0)$, this solution reduces to a particular static Tomimatsu-Sato solution which can be shown to 
be equivalent to the $q-$metric with $q=1$, so that the quadrupole moment is entirely determined by the mass monopole. In the stationary case $(a\neq 0)$, 
the mass quadrupole is $M_2 = -1/4m(m^2-a^2)$ and depends on the rotation parameter and the mass monopole. This indicates that deviations from spherical symmetry are 
due to rotation only and there is no parameter that could be changed in order to modify the deviations. In the case of all the static metrics mentioned above, 
there is always a free quadrupole parameter ($q_2$, $A_2$, $B_2$ or $q$) that is responsible for the deviations. This can be interpreted as an indication that 
metrics with arbitrary quadrupole could describe more general configurations of compact stars. 

Most stationary and axisymmetric solutions in empty space have been obtained by using the solution generating methods mentioned above. Here, we present
the explicit expressions for a stationary metric which can be interpreted as a rotating Erez-Rosen spacetime. This metric has been obtained by applying 
Lie transformations on the Erez-Rosen metric under the condition that the properties of asymptotic flatness and elementary flatness are preserved. 
In prolate spheroidal coordinates, the metric functions can be written as  \cite{quevmas85,quevmas90,quevmas91}
\begin{eqnarray}
f&=&\frac{R}{L} e^{-2q_2 P_2Q_2 }\ , \nonumber\\
\omega&=&-2a-2\sigma\frac{\mathcal M}{R} e^{2q_2 P_2Q_2}\ , \nonumber\\
e^{2\gamma}&=&\frac{1}{4}\left(1+\frac{m }{\sigma}\right)^2\frac{R}{x^2-y^2} e^{2\hat\gamma}\ ,
\end{eqnarray}
where
\begin{eqnarray}
R&=& a_+ a_- + b_+b_-\ , \qquad 
L = a_+^2 + b_+^2\ , \nonumber\\
\quad
{\mathcal M}&=&\alpha x(1-y^2)(e^{2q_2 \delta_+}+e^{2q_2 \delta_-}) a_+ +y(x^2-1)(1-\alpha^2e^{2q_2 (\delta_++\delta_-)})b_+\ , \nonumber\\
\quad
\hat\gamma&=&\frac12(1+q_2 )^2 \ln\frac{x^2-1}{x^2-y^2} + 2q_2 (1-P_2)Q _1 + q_2 ^2(1-P_2) \bigg[ (1+P_2)(Q_1^2-Q_2^2)\nonumber \\
\quad
&&+\frac12(x^2-1)(2Q_2^2 - 3xQ_1Q_2 + 3 Q_0Q_2 - Q_2')\bigg] \ .
\end{eqnarray}
Here $P_l(y)$ and $Q_l(x)$ are Legendre polynomials of the first and second kind, respectively. 
Furthermore
\begin{eqnarray}\quad
a_\pm &=& x( 1-\alpha^2e^{2q_2 (\delta_++\delta_-)})\pm ( 1+\alpha^2e^{2q_2 (\delta_++\delta_-)})\ , \nonumber\\
\quad
b_\pm &=& \alpha y ( e^{2q_2 \delta_+}+e^{2q_2 \delta_-}) \mp \alpha ( e^{2q_2 \delta_+}- e^{2q_2 \delta_-}) \ , \nonumber\\
\quad
\delta_\pm &=& \frac12\ln\frac{(x\pm y)^2}{x^2-1} +\frac32 (1-y^2\mp xy)+\frac{3}{4}[x(1-y^2) \mp y (x^2-1)]\ln\frac{x-1}{x+1}\ ,
\nonumber
\end{eqnarray}
the quantity $\alpha$ being a constant
\be
\label{metquev}
\alpha=\frac{\sigma-m }{a}\ , \qquad \sigma = \sqrt{m ^2-a^2}\ .
\ee

The physical significance of the parameters entering this metric can be established by calculating the 
Geroch-Hansen \cite{ger70a,ger70b,han74} multipole moments 
\begin{equation} 
M_{2k+1} = J_{2k}=0 \ ,  \quad k = 0,1,2,... 
\end{equation} 
\begin{equation} 
M_0 = m \ , \quad M_2 = - ma^2 + \frac{2}{15}q_2 m^3 \left(1-\frac{a^2}{m^2}\right)^{3/2}  \ , ... 
\label{elemm}
\end{equation} 
\begin{equation} 
J_1= ma \ , \quad J_3 = -ma^3 +  \frac{4}{15}q_2 m^3 a \left(1-\frac{a^2}{m^2}\right)^{3/2}  \ , ....
\label{magmm}
\end{equation} 
The vanishing of the odd gravitoelectric ($M_n$) and even gravitomagnetic ($J_n)$ multipole moments is a consequence of the symmetry with 
respect to the equatorial plane $\theta=\pi/2$. It follows from the above expressions  that $m$ is the total mass of the gravitational source, 
$a$ represents the 
specific angular momentum, and $q_2$  is related to the deviation from spherical symmetry. All higher multipole moments can be shown to depend
only on the parameters $m$, $a$, and $q_2$. 
The above solution coincides with the Kerr metric in the limiting case $q_2=0$, and with the Erez-Rosen metric for $a=0$. It also satisfies all
the physical conditions mentioned in the previous section. Therefore, it can be used to describe the exterior field of compact stars.

In the previous section, we presented the $q-$metric as the simplest generalization of the Schwarzschild metric which contains a free quadrupole 
parameter. Therefore, it can be expected that the stationary generalizations of the $q-$metric should also have a simple representation. To show this, 
we apply a particular Lie transformation to the Ernst potential  
\be 
E= \left(\frac{ x-1}{x+1}\right)^{1+q}
\ee
of the $q-$metric in prolate spheroidal coordinates. To obtain the explicit form of the new stationary Ernst potential, we use the solution generating techniques that allow us to generate stationary solutions from a static solution. The procedure involves several differential equations which must be solved under the condition of asymptotic flatness. 
Here, we only present the final expression for the new Ernst potential \cite{tq14}
\be
E= \left(\frac{x-1}{x+1}\right)^q \frac{x-1+(x^2-1)^{-q} d_+} {x+1+ (x^2-1)^{-q} d_- }\ ,
\label{qstar1}
\ee 
where 
\be
d_\pm = \alpha^2 (1\pm x)h_+ h_- + i \alpha [y (h_++h_-)\pm (h_+-h_-)]\ ,
\ee
\be
h_\pm = (x\pm y)^{2q} \ , \quad x= \frac{r}{m} -1 \ , \quad y = \cos\theta \ .
\ee
The new parameter $\alpha$ is introduced by the Lie transformation. As expected, we obtain the $q-$metric in the limiting case $\alpha=0$. 
The behavior of the Ernst potential shows that this new solution is asymptotically flat. The corresponding metric functions corroborate this result. 
Furthermore, the behavior of the new potential near the axis, $y=\pm 1$, shows that the spacetime is free of singularities outside a spatial region determined 
by the radius $x_s=\frac{m}{\sigma}$, which in the case of vanishing $\alpha$, corresponds to the exterior singularity situated at $r_s=2m$. The expression for the Kretschmann scalar shows that the outermost singularity is situated at $x_s=\frac{m}{\sigma}$. Inside this singular hypersurface, several singular structures can appear that depend on the value of $q$ and $\sigma$.

The coordinate invariant multipole moments as defined by Geroch and Hansen \cite{ger70a,ger70b,han74} 
can be found by  using a procedure proposed in \cite{quev90} that allows us 
to perform the computations directly from the Ernst potential. In the  limiting case $q=0$, with  $\alpha=\frac{\sigma-m}{a}$, the  resulting multipoles are
\begin{equation} 
M_{2k+1} = J_{2k}=0 \ ,  \quad k = 0,1,2,... 
\end{equation} 
\begin{equation} 
M_0 = m \ , \quad M_2 = - ma^2   \ , ... 
\end{equation} 
\begin{equation} 
J_1= ma \ , \quad J_3 = -ma^3  \ , ....
\end{equation}
which are exactly the mass $M_n$ and angular $J_n$ multipole moments of the Kerr solution. 
In the general case of arbitrary $q$ parameter, we obtain the following multipole moments 
\be
M_0= m+\sigma q\ ,
\ee
\be
M_2=\frac{7}{3} \,{\sigma}^{3}q-\frac{1}{3}\,{\sigma}^{3}{q}^{3}+m{\sigma}^{2}-m{\sigma}^{2
}{q}^{2}-3\,{m}^{2}\sigma\,q-{m}^{3}\ ,
\ee
\be
J_1= ma + 2a \sigma q\ ,
\ee
\be
J_3 =  -\frac{1}{3}\,a ( -8\,{\sigma}^{3}q+2\,{\sigma}^{3}{q}^{3}-3\,m{\sigma}^{
2}+9\,m{\sigma}^{2}{q}^{2} +12\,{m}^{2}\sigma\,q+3\,{m}^{3})\ .
\ee
The even gravitomagnetic and the odd gravitoelectric multipoles vanish identically 
because the solution is symmetric with respect to the equatorial plane  $y=0$.   Moreover,
higher odd gravitomagnetic and even gravitoelectric multipoles are all linearly dependent since they are completely determined by the parameters 
$m$, $a$, $\sigma$ and $q$. 

In this section, we have seen that there are several exact solutions with quadrupole moment that can be used to describe the exterior field of 
compact stars. This is in accordance with the black hole uniqueness theorems because the presence of the quadrupole invalidates the conditions
under which the theorems have been proved. On the other hand, all the quadrupolar solutions must contain naked singularities, also as a consequence
of the black hole uniqueness theorems. In the case of the stationary $q-$metric and the rotating Erez-Rosen spacetime, we have shown explicitly that
the naked singularities are located inside or on the Schwarzschild radius which in compact stars is always located inside the surface of the star. 
We do not know if this is also true in the case of other quadrupolar metrics mentioned in this section. Suppose, for instance, that a particular 
quadrupolar metric has a singularity at a distance of say $15km$ from the center of a source with a mass of $2M_\odot$. Then, this metric cannot be 
used to represent the exterior field of an isolated neutron star whose radius is about $11.5km$, i.e., the singularity is located outside the surface of the
neutron star where the spacetime should be vacuum. Nevertheless, such a solution can still be a candidate to describe the exterior field, for instance, of a white dwarf 
of mass $1.1M_\odot$ whose radius is of the order of thousand kilometers,  so that the curvature singularity could be located inside the star.

The above discussion is related to the conditions that a general solution must satisfy in order to become physically meaningful. Indeed, if a curvature 
singularity is present, it should be possible to  ``cover'' it by an interior solution that can be matched with an exterior solution across the surface of the
star. In our opinion, the problem of solving the matching conditions in the presence of a physically meaningful interior solution is one of the most important
challenges of modern relativistic astrophysics in general relativity. It is also an important conceptual problem since general relativity, as a theory 
of gravity, should be able to describe physical configurations like compact stars in which the gravitational field plays an important role. We will consider this issue in the next section.

%%%%%%%%%%%%%%%%%%%%%%%%%%%%%%%%%%%%%%%%%%%%%%%%%%%%%%%%%
%%%%%%%%%%%%%%%%%%%%%%%%%%%%%%%%%%%%%%%%%%%%%%%%%%%%%%%%%

\section{Interior quadrupolar metrics }
\label{sec:interior} 

The problem of finding an interior solution for a stationary and axisymmetric spacetime is still open. Even in the case of  vanishing quadrupole, the problem 
is still not completely solved. Indeed, in the case of a perfect fluid with constant energy density, an interior Schwarzschild solution can be obtained analytically, 
but its physical properties do not allow us to use it to describe the interior field of a spherically symmetric compact star because it violates causality, i.e., 
a sound wave propagates inside the star with superluminal velocity. Other spherically symmetric interior solutions are usually non-physical or cannot be matched with the exterior Schwarzschild metric \cite{solutions}. In the case of quadrupolar metrics, the situation is quite similar. The only rigidly rotating perfect-fluid 
solution, containing the Kerr spacetime in the vacuum limit, is the Wahlquist metric \cite{wah68,wah92} which, however,  is characterized by an unphysical 
equation of state $(\rho + 3p = const.)$. Moreover, in the slow rotation approximation, the zero pressure surface corresponds to a prolate ellipsoid rather than an oblate ellipsoid, as expected from a physical point of view. Other solutions with quadrupole represent anisotropic fluids \cite{herjim82,gurgur75,pap01} which,
however, either they do not satisfy the energy conditions \cite{herjim82,gurgur75} or either the boundary surface of zero pressure cannot be fixed because
the hydrostatic pressure cannot be isolated from the other stresses \cite{pap01}. 

All the interior solutions mentioned above have been obtained by analyzing carefully the corresponding field equations and, as we have seen,
 the results are not very satisfactory. In view of this situation, we believe that it is necessary to apply a different approach. We propose to develop 
solution generating techniques for interior spacetimes. Indeed, the discovery of Lie symmetries, B\"acklund transformations and the inverse scattering method 
in the Ernst equations represented a radical change in the search for exterior stationary and axisymmetric solutions.  We believe that a similar approach could be 
useful also in the case of interior solutions.

To illustrate the problem of finding interior solutions, we first consider the case of spherically symmetric spacetimes. To this end, let us consider the following 
line element in spherical coordinates
\be
ds^2= e^{\phi(r)} dt^2 - \frac{dr^2}{1-\frac{2m(r)}{r}} - r^2 (d\theta^2 + \sin^2\theta d\varphi^2) \ .
\ee
We choose a perfect fluid as the physical model for the interior gravitational field. Then, Einstein's equations
\be
R_{\mu\nu} - \frac{1}{2} R g_{\mu\nu} = 8\pi [ (\rho + p) u_\mu u_\nu - p g_{\mu\nu} ] \ 
\ee
reduce to 
\be
\frac{ d \phi}{dr} = \frac{ m + 4\pi r^3 p}{r(r-2m)} \ , \quad m = 4\pi \rho r^2 \ .
\ee
In addition, there is a second order differential equation which is equivalent to the energy-momentum conservation law $T^{\mu\nu}_{\ \ ;\mu} =0$. In this case, 
it  can be written  as the Tolman-Oppenheimer-Volkoff equation 
\be
\frac{dp}{dr}  = - \frac{(p+\rho) (m + 4\pi r^3 p)}{r(r-2m)}\ .
\ee
We see that we have only three equations for determining four unknowns ($\phi$, $m$, $p$, and $\rho$). To close the system of differential equations, it is necessary to impose
an additional condition which is usually taken as the equation of state $p=p(\rho)$. In particular, one can use the barotropic equation of state $p= w\rho$, where $w$ 
is the constant barotropic factor. Many barotropic solutions are known in the literature \cite{solutions} which, however, usually are either not related to realistic equations of state or show a singular behavior at the level of the  pressure or  energy density. To obtain more realistic solutions, we propose to start from 
a physically realistic energy density, for instance. Indeed, suppose that the energy density is given {\it a priori} by the polynomial equation \cite{tqa12}
\be
\rho(r) = \rho_c - c_1 r - c_2 r^2 - c_3 r^3 \ ,
\ee
where $c_1$, $c_2$, and $c_3$ are real constants and $\rho_c$ is the energy density at the center of the body. Then, the mass function can be integrated explicitly 
and we obtain
\be 
m(r) = \frac{\pi}{15} r^3 ( 20 \rho_c - 15 c_1 r - 12 c_2 r^2 - 10 c_3 r^3) \ .
\ee

Clearly, the above particular Ansatz allows us
to obtain a realistic behavior for the energy density, provided the constants are chosen appropriately. For instance, at the surface of the sphere $r=R$ 
we demand that the energy density vanishes, $\rho(r=R)=0$, and so we obtain
\be
\rho_c =  c_1 R + c_2 R^2 + c_3 R^3 \ ,
 \ee
which establishes an algebraic relationship between the free constants. The mass function $m(r)$ is then determined by the free constants only. 
Moreover, we impose the physical condition that the total mass 
\be
M = \int_0^R m(r) dr 
\ee
coincides with the mass of the exterior Schwarzschild metric which implies a boundary condition for the function $\phi(r)$, namely
\be
e^{2\phi(r=R)} = 1 - \frac{2M}{R} \ .
\ee

The procedure consists now in solving the differential equations for $\phi(r)$ and $p(r)$ with the boundary conditions specified above. We did not success in 
finding analytic solutions and, therefore, we integrate the system of differential equations numerically. To this end, it is necessary to impose additional boundary 
conditions as follows. At the center and at the surface of the sphere, the pressure must satisfy the boundary conditions
\be
p_R \equiv p(R) = 0 \ , \quad p(r=0)\equiv p_c < \infty \ .
\ee
Moreover, we demand that the pressure is a well behaved function inside the sphere, i.e., 
\be
0< p(r) < \infty \quad {\rm for} \quad 0 \leq r \leq R \ ,
\ee
which means that the pressure function should be free of singularities inside the sphere.

\begin{figure}
\includegraphics[scale=0.33]{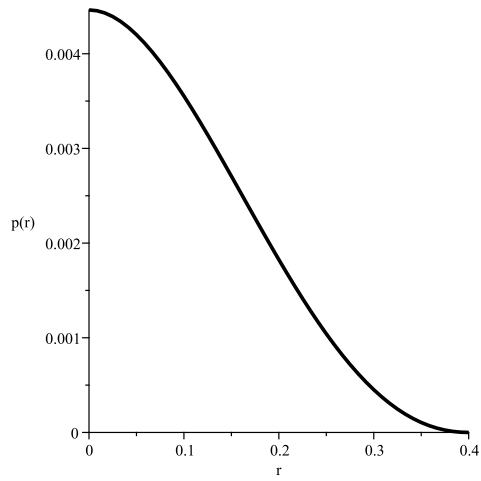}%
\caption{Behavior of the pressure inside a spherically symmetric body of radius $R=0.4$. All the conditions for the pressure to be physically meaningful are satisfied.}
\label{fig1}%
\end{figure}

The method consists now in integrating numerically the 
equations for the total mass $M$ and for the pressure $p$, under the conditions mentioned above. The goal is to find values for the constants $c_1$, $c_2$ and
$c_3$ such that $M$ is positive and $p(r)$ is positive and free of singularities. In fact, it turns out that there are several intervals of values in which all conditions 
are satisfied. The particular simple choice
\be
c_1 = \frac{1}{2} \ ,\quad c_2 = \frac{1}{6}  \ , \quad c_3 = \frac{1}{24} 
\ee
with the particular radius value 
\be
R=0.4
\ee
leads to boundary values
\be
\rho_c = 0.2293\ , \quad 
M=0.002633 \ ,\quad \phi(R) = - 0.0066 \  .
\ee
Then, the integration of the differential equation for the pressure is straightforward. In Fig. \ref{fig1}, we illustrate the behavior of the
pressure. The graphic shows that everywhere inside the sphere, the pressure has a very physical and realistic behavior.
The corresponding function for the energy density shows also a physical behavior as demanded {\it a priori} with the polynomial Ansatz and the chosen values for the 
constants $c_1$, $c_2$ and $c_3$.

\begin{figure}
\includegraphics[scale=0.33]{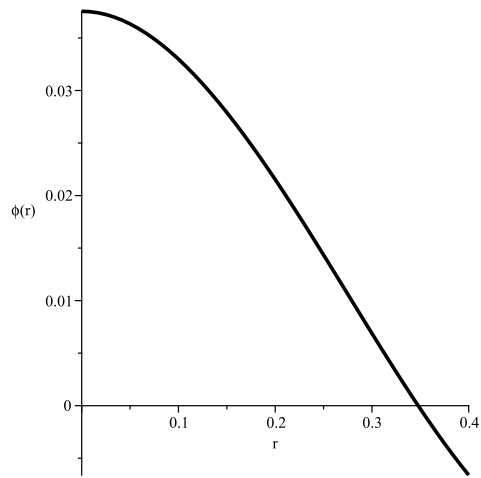}%
\caption{Numerical integration of the metric function $\phi(r)$ for a sphere of radius $R=0.4$ and energy density constants 
$c_1 = \frac{1}{2} \ , c_2 = \frac{1}{6}  \ , c_3 = \frac{1}{24}$.}  
\label{fig2}%
\end{figure}

The differential equation for the function $\phi(r)$ can also be integrated and its behavior is represented in Fig. \ref{fig2}.
It can be seen that this function is well behaved inside the sphere. Moreover, the value at the boundary $R=0.4$, together with the value of the total 
mass, matches exactly the corresponding metric function for the exterior Schwarzschild solution. 

\begin{figure}
\includegraphics[scale=0.33]{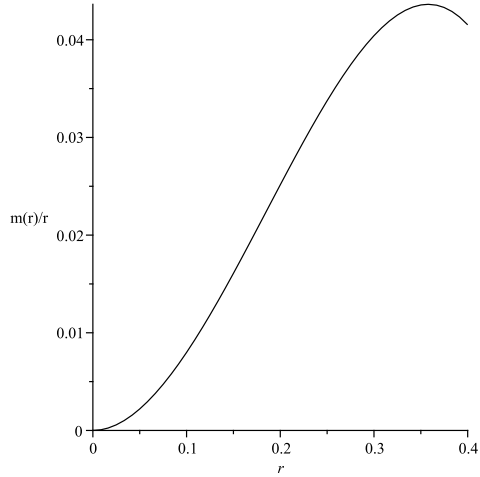}%
\caption{Buchdahl's limit inside the compact sphere of radius $R=0.4$.
The condition $m(r)/r<4/9=0.44$ is satisfied everywhere inside the sphere. }
\label{fig3}%
\end{figure}

An important condition that must be satisfied by any interior solution is the Buchdahl limit \cite{solutions} which, in principle, can be associated
with the Chandrasekhar limit about the maximum mass of compact stars. An analysis of the differential equations that determine the spherically symmetric 
case under consideration here shows that in order to avoid unstable configurations, which could lead to a collapse of the sphere, it is necessary that
the condition $\frac{M}{R} < \frac{4}{9}$ be satisfied. In fact, for a mass-to-radius ratio with $\frac{M}{R} \geq \frac{4}{9}$, the gravitational collapse 
is imminent and the staticity condition of the mass distribution is no longer valid.  So, Buchdahl's limit is an essential requirement for a solution to be
physically meaningful. From the boundary values obtained above, it is easy to see that this requirement is satisfied at the surface of the body. However, it
could be that the behavior of the mass function inside the star violates Buchdahl's limit for a specific value of the radial coordinate, leading to an
internal instability.  To corroborate the stable behavior inside the body, we plot in Fig. \ref{fig3} the behavior of the mass function for all values of
the radial coordinate. 
We can see that inside the sphere the mass-to-radius ratio is everywhere less than the limiting value  $\frac{4}{9}$, indicating that no instabilities can 
occur. This result reinforces the physical interpretation of the numerical solution presented here.

The simple example for a static perfect-fluid sphere as a source of a compact star shows 
that it is possible to find physically meaningful solutions of the interior field equations. 
But it also shows that it is very difficult to integrate analytically the resulting differential equations. We started from a particular polynomial Ansatz for the 
energy density which guarantees a meaningful physical behavior. This allows us to integrate the mass function, but the pressure and the remaining metric function 
$\phi(r)$ cannot be integrated analytically. A numerical analysis seems to be always necessary. For this reason we believe that the standard method of solving directly
the field equations should be complemented by a solution generating technique, similar to the methods used for obtaining exterior solutions.

We now turn back to the study of quadrupolar interior metrics. As mentioned above, all the known solutions are either unphysical or they cannot be 
matched with the exterior Kerr metric. To attack this problem, we propose to consider the mass quadrupole as an additional degree of freedom and 
to analyze the symmetry properties of the field equations in the presence of matter. To begin with, we have considered first the case of static quadrupole
metrics with a perfect fluid as the source of gravity. If we consider the spherically symmetric line element analyzed above, and try to generalize it 
to include the case of axisymmetric fields, it turns out to be convenient to use the following line element \cite{qt15}
\be
ds^2 = e^{2\psi} dt^2 - e^{-2\psi}\bigg[e^{2\gamma}\left(\frac{dr^2}{h} + d\theta^2\right) + \mu^2 d\varphi^2\bigg]\ ,
\label{lel1}
\ee
where $\psi=\psi(r,\theta)$, $\gamma=\gamma(r,\theta)$, $\mu=\mu(r,\theta)$, and $h=h(r)$. A  detailed analysis of the Einstein equations with an energy-momentum 
tensor represented by a perfect fluid shows that the resulting set of differential equations can be split into two systems in a manner which resembles the splitting 
in the case of vacuum spacetimes. Indeed, the main field equations can be written as 
\be
\frac{\mu_{,rr}}{\mu} +  \frac{\mu_{,\theta\theta}}{h\mu} +\frac{h_{,r}\mu_{,r}}{2h\mu} = \frac{16\pi}{h} p  e^{2\gamma - 2\psi}\ ,
\label{mu}
\ee 
\be
\psi_{,rr} + \frac{\psi_{,\theta\theta}}{h} +\left(\frac{h_{,r}}{2h}+\frac{\mu_{,r}}{\mu}\right) \psi_{,r} + \frac{\mu_{,\theta} \psi_{,\theta}}{h\mu} 
= \frac{4\pi}{h}(3  p + \rho)  e^{2\gamma - 2\psi} \ .
\label{psi}
\ee
Moreover, the metric function $\gamma$  is determined by two first order differential equations    
\be
\gamma_{,r} = \frac{1}{h\mu_{,r}^2+\mu_{,\theta}^2}\bigg\{\mu\left[\mu_{,r}\left(h\psi_{,r}^2-\psi_{,\theta}^2\right) +2\mu_{,\theta}\psi_{,\theta}\psi_{,r} +  8\pi\mu_{,r}\bar p\right]
+\mu_{,\theta}\mu_{,r\theta}-\mu_{,r}\mu_{,\theta\theta}  \bigg\}\ ,
\label{gamr}
\ee
\be
\gamma_{,\theta} = \frac{1}{h\mu_{,r}^2+\mu_{,\theta}^2}\bigg\{ \mu\left[\mu_{,\theta}\left(\psi_{,\theta}^2-h\psi_{,r}^2\right) +2 h \mu_{,r}\psi_{,\theta}\psi_{,r}  - 8\pi\mu_{,\theta}\bar p \right] 
+ h\mu_{,r}\mu_{r\theta} +\mu_{,\theta}\mu_{,\theta\theta}\bigg\}\ ,
\label{gamt}
\ee
where 
\be
\bar p = p  e^{2\gamma - 2\psi}\ .
\ee
The equations for $\gamma$ can be integrated by quadratures once the main field equations (\ref{mu}) and (\ref{psi}) are solved, and the pressure $\bar p$ is given {\it a priori} as an independent function.  Notice that if we introduce the differential equations (\ref{mu})-(\ref{gamt}) into the original Einstein equations, a second order differential equation for $\gamma$ is obtained 
\be
\gamma_{,rr}+\frac{\gamma_{,\theta\theta}}{h}+\psi_{,r}^2 + \frac{\psi_{,\theta}^2}{h} + \frac{h_{,r}\gamma_{,r}}{2h}= \frac{8\pi}{h}\bar p \ ,
\label{gam2}
\ee
which must also be satisfied. However, a straightforward computation shows that this equation is identically satisfied if the two first-order 
differential equations (\ref{gamr}) and 
(\ref{gamt}) for $\gamma$ and the conservation equation for the  parameters of the perfect fluid
\be 
p_{,r} = -(\rho+p)\psi_{,r}  \ ,\quad p_{,\theta} = -(\rho+p)\psi_{,\theta} \ ,
\label{claw}
\ee
are satisfied. The conservation equations  resemble the Tolman-Oppenheimer-Volkov relation for  the spherically symmetric case.

We see that the particular choice of the above line element leads to a splitting of the field field equations into two separated sets of equations, and to a generalization of the Tolman-Oppenheimer-Volkov equation for the case of two spatial coordinates.  
This is an important advantage when trying to perform the integration of the main field equations. Indeed, in this manner we found a series of relatively 
simple approximate solutions with non-trivial quadrupole moment. The presentation and physical investigation of those solutions requires several detailed analysis which are beyond the scope of this work, and will be presented elsewhere. A byproduct of such analysis was the discovery of certain symmetries of the field equations 
for a perfect fluid which can be used to generate new solutions from known ones by using the procedure described below.

Suppose that an exact interior solution of Einstein's equations (\ref{mu})-(\ref{gamt}) for the static axisymmetric line element (\ref{lel1}) is  given explicitly by means of the functions 
\be 
h_0 = h_0(r), \ \mu_0=\mu_0(r,\theta), \ \psi_0=\psi_0(r,\theta), \  \gamma_0=\gamma_0(r,\theta),
\label{sol}
\ee
\be 
 \bar p_0 = \bar p_0(r,\theta), \ \bar\rho_0 = \bar\rho_0(r,\theta) \ ,
\ee 
where we have introduced the notation 
\be
\bar p_0 = p_0   e^{2\gamma_0 - 2\psi_0}\ , \quad  \bar \rho_0 = \rho_0   e^{2\gamma_0 - 2\psi_0}\ ,
\ee
and $ p_0$ and $ \rho_0$ are also known functions. Then, for any arbitrary real values of the constant parameter $\delta$, 
a class of new solutions  of the field equations (\ref{mu})-(\ref{gamt}) can be obtained explicitly 
from the functions
\be
h=h_0(r)  ,\  \mu = \mu_0(r,\tilde \theta) , \ \psi = \delta \psi_0(r,\tilde\theta) , 
\label{nsol}
\ee
\be
\ \bar p = \delta \bar p_0(r,\tilde\theta) , \ \bar\rho = \delta \bar\rho_0(r,\tilde\theta), \ \tilde\theta = \frac{\theta}{\sqrt{\delta}} \ ,
\ee
\be
\gamma(r,\tilde\theta)  = \delta^2 \gamma_0(r,\tilde\theta) + (\delta^2 -1 ) \int \frac{\nu_{\tilde\theta}}{h_0+\nu^2} dr
 + 8\pi \delta(1-\delta) \int \frac{ \frac{\mu_0}{\mu_{0,r}} \bar p_0}{h_0+\nu^2} dr + \kappa\ ,
\label{ngam}
\ee
where $\kappa$ is an arbitrary real constant and 
\be 
\nu =\frac{\mu_{0,\tilde\theta}}{\mu_{0,r}} \ .
\ee

To illustrate the application of this solution generating method, let us consider the  spherically symmetric Schwarzschild solution 
which describes the interior field of a perfect-fluid sphere of radius $R$ and total mass $m$. 
The corresponding line element can be written as
\be
ds^2 = \left[\frac{3}{2}f(R) - \frac{1}{2}f(r)\right]^2 dt^2 - \frac{dr^2}{f^2(r)} - r^2(d\theta^2 + \sin^2\theta d\varphi^2)\ ,
\ee
with 
\be
f(r)= \sqrt{1-\frac{2mr^2}{R^3}}\ .
\ee
The physical parameters of the perfect fluid are the constant density $\rho_0$ and the pressure $p_0$, which is a function of the radial coordinate $r$ only 
\be
p_0 = \rho_0 \frac{f(r) - f(R)}{3f(R)-f(r)} \ .
\ee

We now consider the interior Schwarzschild metric as the seed solution (\ref{sol}) for the general transformation (\ref{nsol}). 
A straightforward comparison with the general line element (\ref{lel1}) yields
\be
e^{\psi_0} = \frac{3}{2}f(R) - \frac{1}{2}f(r)\ , \ h_0 = r^2 f^2(r) \ , \ \mu_0 = r \sin\theta \, e^{\psi_0}\ , \ e^{\gamma_0} = r e^{\psi_0} \ .
\label{schint1}
\ee
According to the procedure described above, the new solution can be obtained from Eq.(\ref{schint1}) by multiplying the corresponding metric functions with the new parameter $\delta$. Then, the new line element can be represented as
\be
ds^2 = e^{2\delta \psi_0}   dt^2 -    e^{- 2\delta \psi_0}    
         \left[e^{2\gamma} \left(\frac{dr^2}{r^2f^2(r)} + d\tilde \theta ^2 \right) + r^2 e^{2\psi_0}\sin^2 \tilde \theta  d\varphi^2\right] \ ,
\ee
where the new function $\gamma$ is given by
\be
\gamma  =   \delta^2 \gamma_0  + \int \frac{ (1-\delta^2) +  8\pi\delta (1-\delta)\sin^2\tilde\theta r^2 p_ 0 }
{  r f^2(r) (1+r\psi_{0\,r})\sin^2\tilde \theta + \frac{r}{ 1+r\psi_{0\,r}}\cos^2\tilde\theta } dr
 + \kappa\ ,
\label{ngs}
\ee
with
\be
\psi_{0\,r} = \frac{2mr}{R^3 f(r)[3f(R)-f(r)]}\ .
\ee

Moreover, the physical parameters of the perfect-fluid source are
\be
\rho = \delta \rho_0 e^{2\gamma_0-2\gamma + 2(\delta-1) \psi_0}\ \ , \quad p = \delta p_0 e^{2\gamma_0-2\gamma + 2(\delta-1) \psi_0}\ ,
\label{prho}
\ee
from which the equation of state 
\be
p= \frac{p_0}{\rho_0} \rho
\ee
can be obtained. This is clearly not a barotropic equation of state since the seed pressure $p_0$ depends explicitly on the radial coordinate
$r$. Nevertheless, it can be interpreted as a generalized barotropic equation of state $p=w(r) \rho$. Interestingly, the physical parameters 
of the perfect fluid are axisymmetric, but the equation of state preserves spherical symmetry in the sense that the generalized barotropic factor
depends on the radial coordinate only.

Notice that the new function $\gamma$ depends explicitly  on the new coordinate $\tilde\theta$, in contrast to the seed metric function 
 $\gamma_0$ which depends on the radial coordinate $r$ only. This proves that the new solution is not spherically symmetric, but axisymmetric. 
Notice also that  the density and pressure of the new solution are functions of the angular coordinate too, as expected for an axisymmetric 
mass distribution. It is expected that the obtained deviations from spherical symmetry are related to the quadrupole moment of the perfect fluid;
however, a more detailed investigation is necessary to define an interior quadrupole which should be related to the exterior quadrupole. This is 
a task for future works.

%%%%%%%%%%%%%%%%%%%%%%%%%%%%%%%%%%%%%%%%%%%%%%%%%%%%%%%%%%%%%%%%%%%%%%%%%%%
%%%%%%%%%%%%%%%%%%%%%%%%%%%%%%%%%%%%%%%%%%%%%%%%%%%%%%%%%%%%%%%%%%%%%%%%%%%

\section{Conclusions}
\label{sec:con}

In this work, we presented a review of the problem of describing the interior and exterior gravitational field of compact objects in general relativity, which 
include black holes and compact stars (white dwarfs and neutron stars). 
To take into account rotation and deformation of the mass distribution, we consider stationary and axisymmetric solutions of Einstein's equations 
with quadrupole moment. We formulate the physical conditions which, in our opinion, should be satisfied by a Riemannian manifold in order  to represent the 
interior and exterior gravitational field of compact objects. 

 We review the main static solutions in which the quadrupole is represented by a free parameter. We argue that the $q-$metric 
represents the simplest generalization of the Schwarzschild solution with a quadrupole parameter. We then present a particular generalization of the 
Erez-Rosen metric which includes a rotational parameter, and  reduces to the Kerr metric in absence of the quadrupole parameter. In addition, we present
the Ernst potential of a stationary $q-$metric which turns out to be represented by a quite simple expression. 

We notice that in this review, we limited ourselves to the study of the mass quadrupole as additional parameter only. In general and in more realistic situations, 
it is necessary to consider also the electromagnetic field. Fortunately,  the solution generating techniques have been developed also for Einstein-Maxwell equations 
as well and, therefore, the generalization of the vacuum solutions presented in this review to include electromagnetic multipoles is straightforward.

We argue that the interior counterpart of the exterior Kerr metric cannot be found in general relativity because it is directly related to a curvature singularity 
at which the classical theory breaks down. Probably, a quantum description of gravity is necessary in order to understand the interior field of a black hole.   
In the case of compact stars, however, we argue that general relativity should allow the existence of spacetimes which describe both the interior and exterior 
gravitational field. In view of the precarious situation regarding physically meaningful interior solutions, we propose to study the symmetries of the field equations
in order to develop solution generating techniques. We present a particularly simple method which allows us to generate new static and axisymmetric perfect-fluid solutions from known solutions. 

Summarizing, we propose to apply a different strategy to search for interior physically meaningful solutions of Einstein's equations. Firstly, we propose to include 
the quadrupole as an additional degree of freedom and, secondly, we propose to investigate the symmetry properties of the field equations in the presence of matter.
We intend to follow this strategy in forthcoming works.

\subsection*{Acknowledgements}
This work was  supported by DGAPA-UNAM, Grant No. 113514, Conacyt-Mexico, Grant No. 166391, and 
MES-Kazakhstan, Grant No. 3098/GF4.


\begin{thebibliography}{99}


\bibitem{kerr63} R. P. Kerr, {\it  Gravitational field of a spinning mass as an example of algebraically
special metrics} Phys. Rev. Lett. {\bf 11}, 237 (1963).

\bibitem{heus96} M. Heusler, {\it Black Hole Uniqueness Theorems} (Cambridge University Press, Cambridge, UK, 1996). 


\bibitem{def78} F. de Felice, 
{\it Classical instability of a naked singularity}, 
Nature {\bf 273} (1978) 429.

\bibitem{cal79} M. Calvani and L. Nobili, 
{\it Dressing up a Kerr Naked Singularity}, 
Nuovo Cim. B {\bf 51} (1979) 247.

\bibitem{rud98} W. Rudnicki, 
{\it Cosmic censorship in a Kerr-like scenario}, 
Acta Phys. Pol. {\bf 29} (1998) 981 .



\bibitem{quev90} H. Quevedo, {\it Multipole Moments in General Relativity --Static and 
Stationary Solutions--}, Forts. Physik {\bf 38}, 733--840 (1990).



\bibitem{weyl17} H. Weyl, {\it Zur Gravitationstheorie} Ann. Physik (Germany) {\bf 54}, 117 (1917).



\bibitem{lew32} T. Lewis, {\it Some special solutions to the equations of axially symmetric gravitational
fields}, Proc. Roy. Soc. London A {\bf 136}, 179 (1932).

\bibitem{pap66} A. Papapetrou, {\it Champs gravitationnels stationnaires a symetrie axiale},
Ann. Inst. H. Poincare A {\bf 4}, 83 (1966).

\bibitem{solutions} H. Stephani, D. Kramer, M. MacCallum, C. Hoenselaers, and E. Herlt, 
{\it Exact solutions of Einstein's field equations} (Cambridge University Press, Cambridge, UK, 2003).


\bibitem{diehoe} C. Hoenselaers and W. Dietz, {\it   
Solutions Of Einstein's Equations: Techniques And Results}, Lec. Notes Phys.
{\bf 205}, 1 (1984).
 
\bibitem{ernst} F. J. Ernst, {\it New formulation of the axially symmetric gravitational field problem},
Phys. Rev. {\bf 167}, 1175 (1968).


\bibitem{cnq} J. Cortez, D. Nunez and H. Quevedo, {\it 
Gravitational Fields and Non-linear Sigma Models}, Int. J. Theor. Phys. {\bf 40} 251-260 (2001).




\bibitem{hnq09} 
F. J. Hernandez, F. Nettel and H. Quevedo, 
{\it Gravitational Fields as Generalized String Models} 
Grav. \& Cosm. {\bf 15},  109  (2009).




\bibitem{ger70a} R. P. Geroch, {\it Multipole moments I: Flat space}, J. Math. Phys. {\bf 11}, 1955 (1970).

\bibitem{ger70b} R. P. Geroch, {\it Multipole moments I: Curved space}, J. Math. Phys. {\bf 11}, 2580 (1970).


\bibitem{han74} R. O. Hansen, {\it Multipole moments of stationary space-times}, J. Math. Phys. {\bf 15}, 46 (1974).



\bibitem{erro59} G. Erez and N. Rosen, {\it The gravitational field of a particle possessing a quadrupole moment.}
Bull. Res. Counc. Israel {\bf 8}, 47 (1959).


\bibitem{gm85} T. I. Gutsunaev  and V. S. Manko, {\it On the gravitational field of a mass possessing
a multipole moment.} Gen. Rel. Grav. {\bf 17}, 1025 (1985). 

\bibitem{quev87} H. Quevedo, {\it On the Exterior Gravitational Field of a Mass with a 
Multipole  Moment}, Gen. Rel. Grav. {\bf 19}, 1013--1023 (1987).



\bibitem{hm94} J. L. Hern\'andez-Pastora and J. Mart\'\i , {\it Monopole-quadrupole static axisymmetric solutions of 
Einstein field equations}, Gen. Rel. Grav. {\bf 26}, 877 (1994).

\bibitem{fs16} F. Frutos-Alfaro and M. Soffel, 
{\it Multipole moments of the generalized Quevedo-Mashhoon metric}, (2016), arXiv:1606.07173 [gr-qc].


\bibitem{quev11a} H. Quevedo, {\it Mass quadrupole as a source of naked singularities}
Int. J. Mod. Phys. D {\bf 20} 1779 (2011). 


\bibitem{zip66} D. M. Zipoy, 
{\it Topology of some spheroidal metrics.}
 J. Math. Phys. {\bf 7}, 1137 (1966).

\bibitem{voo70} B. Voorhees, 
{\it Static axially symmetric gravitational fields.}
 Phys. Rev. D {\bf 2}, 2119 (1970).



\bibitem{mala04} D. Malafarina, 
{\it Physical properties of the sources of the Gamma metric}, 
Conf. Proc. C0405132, 273  (2004).



\bibitem{ts73} A. Tomimatsu and H. Sato, {\it New series of exact solutions for gravitational
fields of spinning masses} Prog. Theor. Phys. {\bf 50}, 95 (1973). 



\bibitem{bggqt16} K. Boshkayev, E. Gasperin, A.C. Guti\'errez-Pi\~neres, H. Quevedo, and S. Toktarbay, 
{\it Motion of test particles in the field of a naked singularity}, 
Phys. Rev. D {\bf 93}, 024024 (2016).


\bibitem{sib91} N. R. Sibgatullin, {\it Oscillations and Waves in Strong Gravitational and Electromagnetic Fields},
(Springer-Verlag, New York, 1991). 
 
\bibitem{mms00} V. S. Manko, E. W. Mielke, and J. D. Sanabria-G\'omez, 
{\it Exact solution for the exterior field of a rotating neutron star},
Phys. Rev. D {\bf 61}, 081501 (2000).


\bibitem{prs06} L. A. Pach\'on, J. A. Rueda, and J. D. Sanabria-G\'omez, 
{\it Realistic exact solution for the exterior field of a rotating neutron star},
Phys. Rev. D {\bf 73}, 104038 (2006).




\bibitem{quevmas85}
H. Quevedo and B. Mashhoon, {\it Exterior Gravitational Field of a Rotating Deformed Mass}, 
Phys. Lett. A {\bf 109}, 13--18 (1985).

\bibitem{quevmas90} H. Quevedo and B. Mashhoon, {\it Exterior  Gravitational Field of a Charged Rotating Mass 
with Arbitrary Quadrupole Moment}, Phys. Lett. A {\bf 148}, 149--153 (1990).




\bibitem{quevmas91} H. Quevedo   and     B. Mashhoon, {\it Generalization   of   Kerr   Spacetime},
Phys. Rev. D {\bf 43}, 3902--3906 (1991). 






\bibitem{tq14} S. Toktarbay and H. Quevedo, 
{\it  A Stationary q-metric}, 
Grav. \& Cosm. {\bf 20}, 252 (2014).

\bibitem{wah68} H.  Wahlquist, 
{\it Interior solution for a finite rotating body of perfect fluid},
Phys. Rev. {\bf 172}, 1291 (1968).

\bibitem{wah92} H. Wahlquist, 
{\it The problem of exact interior solutions for rotating rigid bodies in general relativity}, 
J. Math. Phys. {\bf 33}, 304 (1992); Erratum {\bf 33}, 3255.

\bibitem{herjim82} L. Herrera and J. Jim\'enez,
{\it The complexification of a non rotating sphere: An extension of the Newman-Janis algorithm}, 
J. Math. Phys. {\bf 23}, 2339 (1982).

\bibitem{gurgur75} M. G\"urses and F. G\"ursey,
{\it Lorentz covariant treatment of the Kerr-Schild geometry},
J. Math. Phys. {\bf 16}, 2385 (1975).

\bibitem{pap01} T. Papakostas, 
{\it Anisotropic fluids in the case of stationary and axisymmetric spaces of general relativity},
Int. J. Mod. Phys. D {\bf 10}, 869 (2001).

\bibitem{tqa12} S. Toktarbay, H. Quevedo and M. Abishev,
{\it Interior solutions of Einstein's equations},
Atomic Physics and Elementary Particles {\bf 5}, 22 (2012); in Russian.  



\bibitem{qt15} H. Quevedo and S. Toktarbay, {\it Generating static perfect-fluid solutions of Einstein's equations.}
J. Math. Phys. {\bf 56}, 052502 (2015).

\end{thebibliography}
\end{document}